\documentclass[
twocolumn,
amsmath,amssymb,aps,pre
]{revtex4-2}

\usepackage[utf8]{inputenc}
\usepackage{times}

\usepackage{xcolor}

\usepackage{graphicx}
\usepackage{amsmath,amssymb}
\usepackage{gensymb,lipsum}
\usepackage{multirow}
\usepackage{verbatim}
\usepackage{dcolumn}
\usepackage{bm}

 \usepackage{scalerel}

\newcommand{\stackrelsp}[2]{%
\ThisStyle{%
\mathrel{\vbox{\offinterlineskip
\ialign{%
\hfil##\hfil\cr
$\SavedStyle#1$\cr
\noalign{\kern-0.2pt}
$\SavedStyle#2$\cr
}%
}}%
}%
}

\begin{document}
\preprint{APS/123-QED}

\title{Thermodynamic Langevin Equations}%




\author{Amilcare Porporato}
\email{aporpora@princeton.edu}
\affiliation{Department of Civil and Environmental Engineering and High Meadows Environmental Institute, Princeton University,
Princeton, New Jersey 08540, USA}
\author{Salvatore Calabrese}
\email{Salvatore.Calabrese@ag.tamu.edu}
\affiliation{Department of Biological and Agricultural Engineering, Texas A \& M University,
College Station, TX, USA}
\author{Lamberto Rondoni}
\email{lamberto.rondoni@polito.it}
\affiliation{Dipartimento di Scienze Matematiche, Politecnico di Torino, Corso Duca degli Abruzzi 24, Turin, Italy}
\affiliation{INFN, Sezione di Torino, Via P. Giuria 1, 10125 Turin, Italy}
\affiliation{ORCID: 0000-0002-4223-6279}

\date{\today}

\begin{abstract}
The physical significance of the stochastic processes associated to the generalized Gibbs ensembles is scrutinized here with special attention to the thermodynamic fluctuations of small systems. The contact with the environment produces an interaction entropy, which controls the distribution of fluctuations and allows writing the generalized Gibbs ensembles for macrostates in potential form. This naturally yields exact nonlinear thermodynamic Langevin equations (TLEs) for such variables, with drift expressed in terms of entropic forces. The analysis of the canonical ensemble for an ideal monoatomic gas and the related TLEs show that introducing currents leads to nonequilibrium heat transfer conditions with interesting bounds on entropy production but with no obvious thermodynamic limit. For a colloidal particle under constant force, the TLEs for macroscopic variables are different from those for the microscopic position, typically used in the so-called stochastic thermodynamics; while TLEs are consistent with the fundamental equation obtained from the Hamiltonian, stochastic thermodynamics requires isothermal conditions and entropy proportional to position.
 \end{abstract}


\maketitle

\section{Introduction}

The advent of nanotechnologies compels us to take fluctuations seriously, from both a theoretical and a practical point of view. Much work has been devoted to formulating a consistent theory for small systems \cite{miller1979heat,hill2001different,dobruskin2006size,inbook,kato2001breaking,SekimotoBook,jarzynski2012equalities,bedeaux2023nanothermodynamics,morgado2014thermostatistics,coasne2014pressure,aslyamov2014some,ciliberto2017experiments,puglisi2018,daivis2018challenges,squarcini2022spectral,maffioli2022slip,liu2021frontiers,khadem2022stochastic}.
This endeavour has been
facilitated by the generality of the thermodynamic formalism (e.g., \cite{He95,jaynes1957information,ruelle2004thermodynamic,daubechies1994thermodynamic,vul2017feigenbaum,lecomte2007thermodynamic}) while constrained by the need of a seamless connection to classical thermodynamics. Matter at mesoscopic scales often features properties that are not fully understood from a thermodynamic viewpoint, such as blurred states of aggregation and anomalous transport.

Fluctuating thermodynamics dates back to Gibbs \cite{gibbs1928collected}, who introduced the index of probability, and Einstein \cite{einstein1910theorie}, who obtained fluctuation probabilities by inverting  Boltzmann's formula. Well-known subsequent developments are due to Callen, Greene, Kubo, and many others  \cite{callen1998thermodynamics,landau2013statistical,greene1951formalism,tisza1963statistical,mishin2015thermodynamic}. Recently, interest has been revived by the hope to find a window into the elusive world of small systems out of equilibrium. It is virtually impossible to do justice to this burgeoning literature, ranging from nanothermodynamics \cite{hill2001different} and the fluctuation theorem \cite{evans2002fluctuation,marconi2008fluctuation,jarzynski2017stochastic} to stochastic thermodynamics \cite{seifert2012stochastic}. 

With this premise, it appeared useful to revisit the connections between classical thermodynamics and stochastic thermodynamics. We started by following Greene and Callen \cite{greene1951formalism}, outlining the key steps leading to the generalized Gibbs ensembles for the macrostates from the classical thermodynamics formalism (Sec. II). From the fact that the availability and the related entropy generation serve as potential functions for the probability distributions of thermodynamic fluctuations (Sec. III), it emerges that the Gibbs ensembles can be interpreted  as exact solutions of nonlinear thermodynamic Langevin equations (TLEs) for macrostates, based on entropic forces and subject to natural boundaries (i.e., in detailed balance; \cite{gardiner1985handbook,van1992stochastic}). 
This formally opens up the possibility of analyzing thermodynamic fluctuations in nonequilibrium conditions (i.e., with probability currents due to transient or special boundary conditions) and of exploring their connection to actual nonequilibrium thermodynamic processes.

The TLEs presented here appear to be novel. Discussion of their physical meaning and usefulness naturally  involves elucidating the extent to which the asymptotic formulations behind the ensembles can be pushed to describe deviations of small systems outside of the thermodynamic limit, where different ensembles \cite{suzen2009ensemble} can no longer be regarded as equivalent \cite{tisza1963statistical,gallavotti1999statistical}. We also focus on questions related to the differences between fluctuations calculated from either macroscopic or microscopic quantities and whether the related entropies and energies are connected to useful physically measurable quantities, such as work and dissipation. Indeed, clarifying the connections between the theories of fluctuations and classical thermodynamics is crucial for a thermodynamic theory of small system to have physical meaning and practical implications. 

For simplicity, we begin our discussion to simple systems, {\it sensu} thermodynamics (i.e., homogeneous and isotropic systems, without surface effects and added electric, magnetic, and gravitational fields). After the general considerations of Sec. II-IV, in Sec. V we focus our attention on the case of monoatomic ideal gas, where analytical considerations allow some more explicit conclusions. Sec. VI presents the case of a colloidal particle, which  is important also because of its prototypical role in stochastic thermodynamics \cite{seifert2012stochastic}. Chemical reactions, for which Gibbs' thermodynamic formalism naturally includes differences in affinities and therefore entropy production, 
will be considered in future contributions.

\section{Potential Form of the Gibbs Ensembles}

\subsection{Thermodynamic Formalism}

The generalized Gibbs ensembles rely on the thermodynamic formalism, whereby equilibrium macroscopic systems are fully described by the fundamental equation (in the entropy representation) \cite{callen1998thermodynamics},  
\begin{equation}
    S=S(\bf{X}),
\label{eq:fund}
\end{equation}
relating the entropy $S$ to the extensive variables ${\bf X}$. For a `simple system', the extensive variables are: internal energy, volume, and number of moles, respectively $U$, $V$, and $N$ (the number of particles is $\tilde{N}= N_A N$, where $N_A$ is Avogadro's number); regardless of the type of thermodynamic system, the first of such variables is always the internal energy, i.e., $X_1=U$ \cite{callen1998thermodynamics}. In the entropy representation, the vector of intensive quantities is the gradient of the fundamental equation,
\begin{equation}
    {\bf y}={\bf \nabla}_{\bf X} S.
\label{eq:intensive}
\end{equation}
For a simple system, these are $1/T$, $-p/T$, $\mu/T$, where $T$ is temperature, $p$ is pressure, and $\mu$ is the chemical potential.

The entropy that would be 
produced by dissipating the maximum work obtainable \cite{callen1998thermodynamics} in a quasistatic equilibration of a system in a state $\hat{{\bf X}}$ to a reference state ${\bf X}$, in equilibrium with the environment characterized by intensive quantities ${\bf y}$, is given by \cite{callen1998thermodynamics,landau2013statistical}
\begin{equation}
    \Sigma(\hat{{\bf X}},{\bf y})={\bf y}\cdot (\hat{{\bf X}}-{\bf X})-(\hat{S}-S)=({\bf y}\cdot \hat{{\bf X}}-\hat{S})+\Phi({\bf y}),
    \label{eq:Sigma}
\end{equation}
where the free entropy, or generalized Massieau function, is defined by
\begin{equation}
    \Phi({\bf y})=S({\bf X}({\bf y}))-{\bf y}\cdot{\bf X}({\bf y}).
\label{eq:massieu}
\end{equation}
Here we have stressed that the Legendre transform makes the old variable {\bf X}
depend on the new variable {\bf y}.
The quantity $\Sigma$, corresponding to the minimum  dissipation, has the dimensions of an entropy associated with the variation of the state from $\hat{\bf X}$ to a reference state ${\bf X}$.
In turn, $\Phi$ is the Legendre transform of the entropy with respect to the intensive variables being varied during the thermodynamic transformation from $\hat{{\bf X}}$ to ${\bf X}$. Thus, for a Legendre transform with respect to $1/T$, $\Phi$ is the Massieu function (i.e., Helmholtz free energy divided by temperature), while for a Legendre transform with respect to both $1/T$ and $-p/T$, $\Phi$ is the Planck function (i.e., Gibbs free energy divided by temperature). For simplicity of notation, if a quantity is held fixed at the reference state (i.e., $X_i=\hat{X}_i$), we assume tacitly that the Legendre transform is not carried out with respect to that variable. 

\begin{figure}
\begin{center}
\includegraphics[width=250pt]{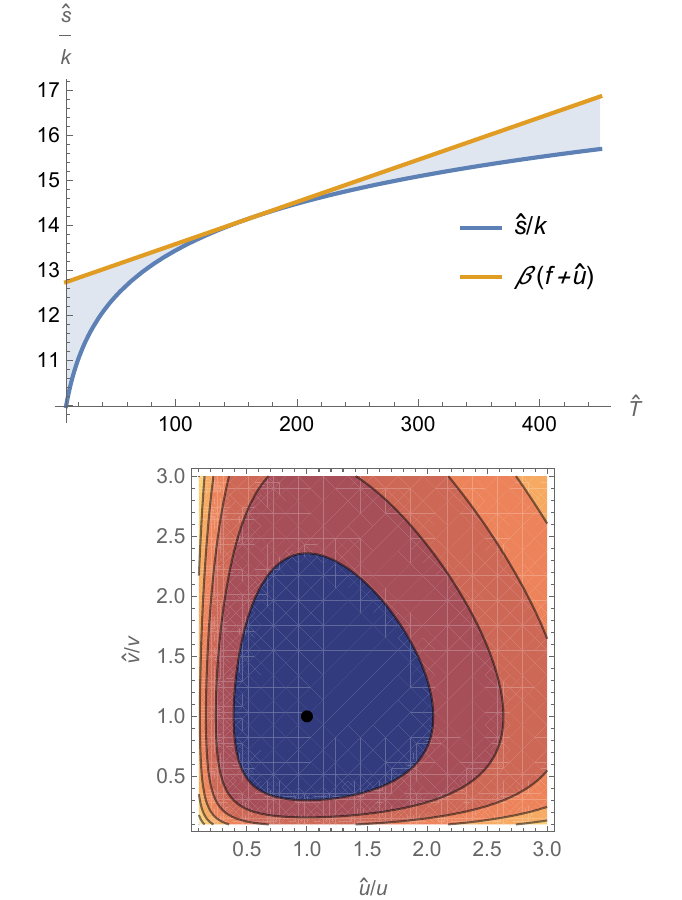}
\end{center}
\caption{Top: Dimensionless particle specific entropy and potential entropy generation $\Sigma$ (shaded area) for changes in temperature ($\hat{T}=2/3\hat{u}$ around equilibrium ($T=2/3 u=160 K$).  Bottom: potential entropy generation $\Sigma$ as a function of both internal energy and volume around equilibrium (black dot) for a monoatomic ideal gas (helium).}\label{fig:entrgen}
\end{figure}

By the Stodola-Guoy theorem \cite{bejan2013entropy}, the availability \cite{kestin1979course}, or maximum work obtainable by reversible transformations, going from $\hat{{\bf X}}$ to ${\bf X}$ \cite{callen1998thermodynamics} is given by 
\begin{equation}
    R(\hat{{\bf X}},{\bf y})=T \Sigma (\hat{{\bf X}},{\bf y}) \ge 0,
    \label{eq:avail}
\end{equation}
where $T$ is the temperature of the reference state. Formally, the availability is a convex surface, which defines a contact structure \cite{mrugala1991contact}; it is obviously zero when the two states are the same ($\hat{S}=S$), namely when ${\bf y}\cdot \hat{{\bf X}}-\hat{S}$ becomes the Legendre transform $\rightarrow \Phi({\bf y})$ of $S({\bf X})$ with respect to ${\bf X}$ (Eq. (\ref{eq:massieu})).
Fig. \ref{fig:entrgen} (top panel) shows the potential entropy generation $\Sigma$ (associated to the minimum dissipation) as a shaded area, see Eq. (\ref{eq:Sigma}), for a monoatomic ideal gas at constant volume and number of moles (i.e., conditions similar to the canonical ensemble), calculated using the Sackur-Tetrode equation for a monoatomic ideal gas, 
\begin{eqnarray}
s=k \ln \left(\left(u\right)^{\frac{3}{2}}v\right)+s_0,
\label{eq:Sackurnorm}
\end{eqnarray}
where $s=S/\tilde{N}$, $v=V/\tilde{N}$, $s_0=+\frac{3}{2}k \left(\frac{5}{3}+\ln \frac{4 \pi m}{3 h^2} \right)$, with $m$ the particle mass, and $h$ the Planck constant.
The bottom panel shows $\Sigma$ as a function of internal energy and volume (corresponding to the conditions of a pT ensemble), as a convex surface with a minimum at the reference equilibrium point. One should recall that (\ref{eq:Sackurnorm}) is derived in the canonical ensemble, making use of the Stirling approximation. Therefore, its use in the case of small $\tilde N$ requires some care, as will be discussed in the following.

\subsection{Generalized Gibbs Ensembles for Macrostates}

We begin, following Khinchin \cite{aleksandr1949mathematical} and Green and Callen \cite{greene1951formalism,callen1998thermodynamics}, by considering an isolated system in equilibrium with fixed values ${\bf X}$ of the extensive quantities. Thus, on the microscopic level, the corresponding mechanical quantities vary within such small intervals that they appear infinitesimal and correspond to single values on the macroscopic scale.
The fundamental equation can then be obtained by the the Boltzmann postulate,
\begin{equation}
    S({\bf X})=k\ln \mathcal{W}({\bf X})=k \ln \frac{\Omega({\bf X})}{\Omega_0},
    \label{eq:postul}
\end{equation} 
where $k$ is the Boltzmann constant, 
and $\mathcal{W}({\bf X})$  is the number of distinguishably distinct microstates of the macrostate ${\bf X}$. These  occupy a volume $d \mathcal{V}=dX_1 dX_2 ... dX_n$ to which 
the macrostate ${\bf X}$ belongs. 
The second equality in (\ref{eq:postul}) allows us to pass from a discrete to a continuous description, considering a density of microstates in the given volume $d \mathcal{V}$, $\Omega({\bf X})=\mathcal{W}({\bf X})/d\mathcal{V}$, compared to the unitary density, i.e., $\Omega_0=1/d\mathcal{V}$ \cite{greene1951formalism,aleksandr1949mathematical}.

The contact with an environment, characterized by the state variable ${\bf y}$, 
allows the previously fixed extensive quantities to fluctuate and explore different states, each concerning a different value of the extensive quantity ${\hat{\bf X}}$, where the hat `$\hat{\cdot }$' indicates fluctuating variables. Writing the canonical occupation probability of the $i$-th microstate, 
fluctuating in a mescoscopic state $\hat{{\bf X}}$, as \cite{greene1951formalism}
\begin{equation}
   f_i(\hat{{\bf X}})=
e^{-(\Phi({\bf y})+{\bf y}\cdot \hat{{\bf X}})/k},
    \label{eq:canmicro}
\end{equation}
the probability density function (pdf) of the macrostates is obtained multiplying such a probability by the density of states, 
\begin{equation}
    p_{\bf y}(\hat{{\bf X)}}  =\Omega(\hat{{\bf X}}) f_i(\hat {\bf X}), 
\label{eq:micromacro}
\end{equation} 
that is 
\begin{equation}
    p_{\bf y}(\hat{{\bf X)}} d\mathcal{V} =\mathcal{W}(\hat{{\bf X}}) f_i(\hat {\bf X}). 
\label{eq:micromacro1}
\end{equation} 
Adopting the Boltzmann formula \eqref{eq:postul} for a given ${\hat{\bf X}}$, we get the generalized Gibbs ensembles for the macrostates \cite{greene1951formalism}
\begin{equation}
    p_{\bf y}(\hat{{\bf X}})
=
\Omega_0 \, 
 e^{-(\Phi({\bf y})+{\bf y}\cdot \hat{{\bf X}}-\hat{S}(\hat{{\bf X}}))/k}.
    \label{eq:ens}
\end{equation}
In Grene and Callen's terminology, such joint distributions are canonical with respect to the extensive quantities that fluctuate, $\hat{{\bf X}}$, and microcanonical with respect to the extensive quantities, ${\bf X}$, that are fixed and therefore not Legendre transformed \cite{greene1951formalism}.
As noted in \cite{greene1951formalism}, using the micronanonical Boltzmann formula \eqref{eq:postul} is legitimate in the present canonical formalism, because the entropy $S({\bf X})$ is a function of the state $\bf X$, and that state corresponds to $\cal W({\bf X})$  microstates, whether $\bf X$
is fixed or allowed to vary in time.

Now, comparing equations (\ref{eq:Sigma}) and (\ref{eq:avail}) with the expression of the generalized ensembles (\ref{eq:ens}), allows us to write
\begin{equation}
    p_{\bf y}(\hat{{\bf X}}) =
\Omega_0 e^{-\Sigma(\hat{{\bf X}},{\bf y})/k}.
    \label{eq:enspot}
\end{equation}
Based on the definition of the intensive quantities (\ref{eq:intensive}), the minimum entropy generation $\Sigma$ (see Eq. \ref{eq:Sigma}) appears as the result of an integration of the entropic forces, which further allows us to write 
\begin{eqnarray}
    p_{\bf y}(\hat{{\bf X}})
&=&
\Omega_0 \,
e^{-\int_{\hat{{\bf X}}}d \hat{\bf X}\cdot({\bf y}-\nabla_{\hat{{\bf X}}}\hat{S})/k} \\
&=&\Omega_0 \, 
e^{-\int_{\hat{{\bf X}}}d \hat{\bf X} \cdot ({\bf y}-\hat{{\bf y}})/k}
    \label{eq:enspot1}
\end{eqnarray}
This shows that $\Sigma$ acts as a potential function for the generalized ensembles written for the macroscopic variables. In term of the availability,  $p_{\bf y}(\hat{{\bf X}}) = \Omega_0  e^{-\beta R(\hat{{\bf X}})}$ 
is reminiscent of the well known form for the phase-space variables, $p(\hat{{\bf z}})=Ze^{-\beta \mathcal{H}(\hat{{\bf z}})}$, where $\mathcal{H}$ is the Hamiltonian, with the key difference that the Hamiltonian in the latter is not a potential. 

\begin{figure}
\begin{center}
\includegraphics[width=265pt]{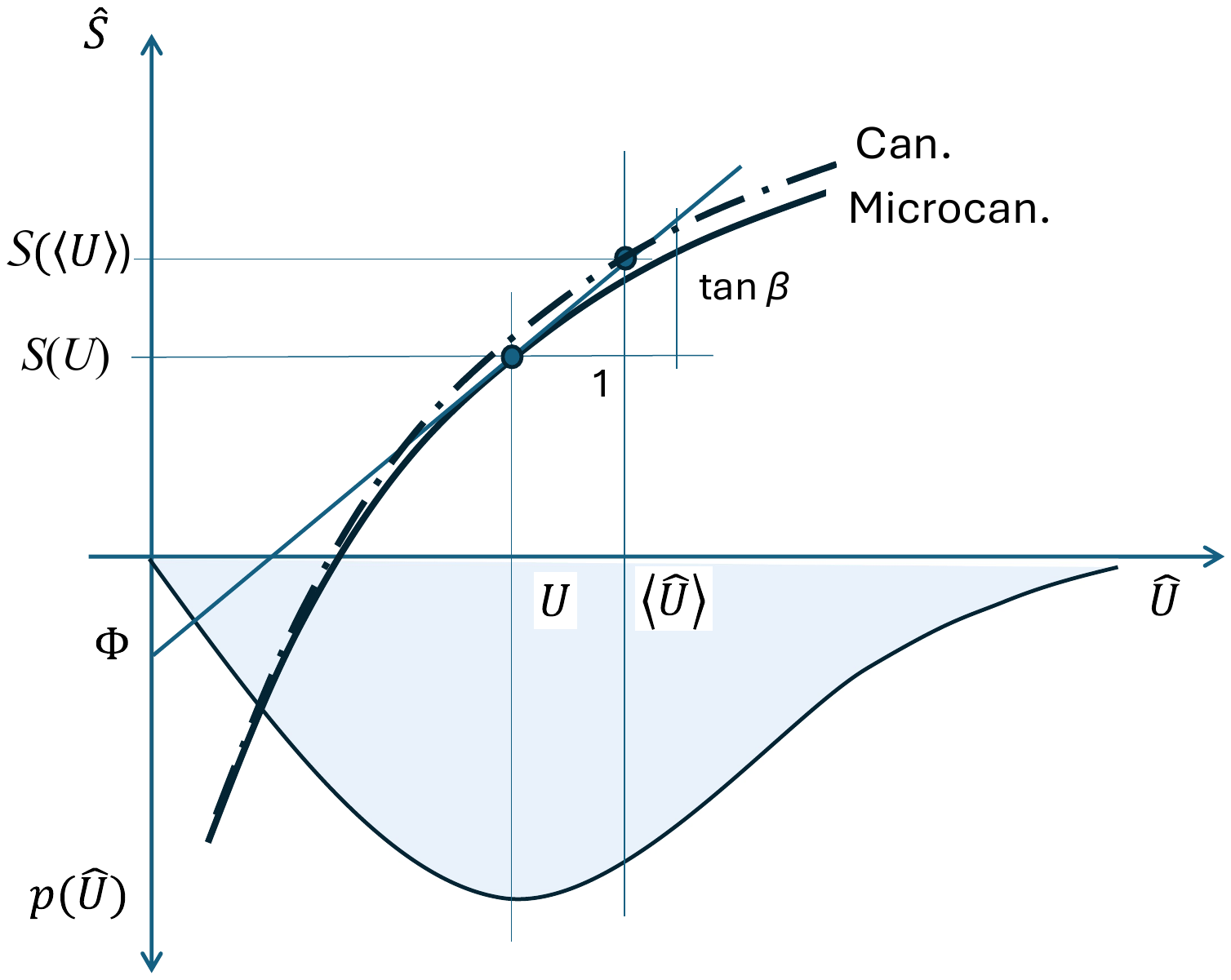}
\end{center}
\caption{For small systems, sketch of the canonical distribution and related thermodynamic fundamental equations, related to the mode (microcanonical) and the mean (canonical).} \label{fig:modemean}
\end{figure}

The fluctuating entropy
\begin{equation}
- k \ln \cfrac{p_{\bf y}(\hat {\bf X})}{\Omega_0}=\Phi({\bf y})+{\bf y}\cdot \hat{{\bf X}}-\hat{S}(\hat{{\bf X}}) \\
=\Sigma(\hat{{\bf X}}\rangle) \ge 0
\label{eq:indexprob}
\end{equation}
is the entropy generated when the canonical system in the specific state $\hat{{\bf X}}$ is brought quasistatically to the state ${\bf X}$ of the original microcanonical ensemble (which is also the mode of the canonical distribution). Ensemble averaging yields the canonical entropy
\begin{equation}
    \mathcal{S}=\langle \Sigma(\hat{{\bf X}}\rangle)  \rangle+\langle \hat{S}\rangle=\Phi({\bf y})+{\bf y}\cdot \langle \hat{{\bf X}}\rangle,
\end{equation}
which reads as a Legendre transform as in Eq. (\ref{eq:massieu}) for the averages and thus implies a fundamental equation 
$\mathcal{S}=\mathcal{S}(\langle \hat{{\bf X}}\rangle)$.
The same value of the canonical entropy is obtained considering the negative of the Shannon entropy of the microstate distribution (\ref{eq:canmicro}).
For small systems, the difference between the entropy of the canonical ensemble and the corresponding microcanonical entropy $S$ is due to the asymmetric fluctuations of the thermodynamics state variables (e.g., $\langle \hat{{\bf X}}\rangle>{\bf X}$). This average effect
of the interaction with the environment due to an increase in the number of the allowed microstates,
can also be described in terms of an interaction energy
\begin{equation}
\mathcal{R}=\Psi+\langle R\rangle=T(\Psi+\langle \Sigma \rangle),    
\end{equation}
as the minimum work necessary to separate the system from the environment (e.g., \cite{delle2017partitioning}); see Fig \ref{fig:modemean}. Such differences become negligible in the thermodynamic limit.

\section{Thermodynamic Langevin Equations} \label{sec:three}

Because of their form (\ref{eq:enspot}), the generalized ensembles can be interpreted as the steady state pdf of a stochastic diffusion process for the macroscopic thermodynamic variables \cite{gardiner1985handbook,van1992stochastic}.
The corresponding thermodynamic Langevin equations (TLEs) have the exact form 
 \begin{equation}
     \alpha d\hat{{\bf X}}=({\bf y}-\hat{{\bf y}})/k \, dt+d{\bf \hat{W}},
     \label{eq:TLE}
 \end{equation}
where $\alpha$ is an arbitrary constant which sets the timescale and takes care of the dimensions, $\hat{{\bf y}}={\bf \nabla}_{\hat{\bf X}} \hat{S}$, and the noise is a multidimensional Wiener process.  The nonlinear drift vector appears in the form of the generalized thermodynamic force. 

In spite of the extensive literature on Langevin equations in statistical mechanics, we are not aware of a previous presentation of the TLEs (\ref{eq:TLE}). In the Mori-
Zwanzig method \cite{zwanzig1973nonlinear,zwanzig2001nonequilibrium} for multiparticle systems, a set of Langevin equations originate from suitable projections of the fast degrees of freedom, while in stochastic thermodynamics Langevin equations  typically come from the stochastic dynamics of configuration degrees of freedom \cite{seifert2012stochastic}) (see Sec. VI for the prototypical case of a colloidal particle)  or from a continuous diffusive limit of chemical master equations \cite{rao2016nonequilibrium}. Here, instead, the TLEs for the macroscopic variable emerge as an exact consequence of the potential form of the potential 
minimum entropy generation $\Sigma$ (or similarly the availability), as the argument of the exponential of the generalized Gibbs ensembles (\ref{eq:enspot}). The nonlinear drift, expressed 
in terms of entropic forces, directly links the stochastic process to the underlying classical thermodynamic formulation. Thus, a part from assuming a continuous trajectory and Markovian dynamics (i.e., diffusion), all the other underlying assumptions coincide with those made in the formulation of the ensembles themselves. The diffusion approximation and the noise intensity set the timescales of the system response. It should be clear that other dynamic models are also compatible with a steady state solution in the form (\ref{eq:ens}).

Because of the potential condition, the steady state solution of the TLEs with natural boundary conditions does not have circulation currents and obeys detailed balance \cite{gardiner1985handbook}, as it is appropriate for a thermodynamic equilibrium condition. However, either because of non stationary conditions, due to specific initial conditions, or because of changes in the environmental parameters (i.e., with time
dependent intensive quantities, ${\bf y}(t)$), or through open boundary conditions, probability currents can be originated. Assuming that the resulting  stochastic processes indeed describe thermodynamic fluctuations of small systems out of equilibrium, then in principle they should also be associated with irreversibilities and entropy generation. One can imagine, for example, to start the stochastic process from an initial condition corresponding to a microcanonical system and let the system evolve toward its canonical distribution at the same temperature. One would then observe an entropy generation, 
presumably related to the energy necessary to isolate the system. Alternatively, a protocol of finite time adiabatic and isothermal compressions and expansions could be devised to reproduce the stochastic dynamics of a finite-time Carnot cycle; or finally, a drift with an external force pulling a colloidal particle could also be used to induce a probability current in nonequilibium conditions. Undoubtedly, such nonequilibrium extensions of the TLEs have tantalizing thermodynamic connotations; however, whether the resulting systems have actual connections to thermodynamic transformations in nonequilibrium conditions should be scrutinized with the aid of specific models for which detailed calculations are possible.

\section{Ideal Gas Ensembles}

The TPN-ensemble describes systems in contact with a reservoir at constant temperature $T$ and pressure $p$, with fixed number of particles,
\begin{equation}
p_{_{{\rm TPN}}}(\hat{U},\hat{V};\beta,\eta,N)=\Omega_0e^{\beta G(\beta,\eta,N)-\beta \hat{U}-\eta \hat{V}+S(\hat{U},\hat{V},N)/k}.
\label{eq:PTNideal}
\end{equation}
Both internal energy $\hat{U}$ and volume $\hat{V}$ are random variables because the system can exchange heat and work with the reservoir; the Planck function, $\Phi=-G(T,p,N)/T$ is the negative of the Gibbs free energy divided by temperature, while the intensive quantities, normalized by the Boltzmann constant, are $\beta=\frac{1}{kT}$ and $\eta=\frac{p}{kT}$. For a monoatomic ideal gas, the Sakur-Tetrode equation (\ref{eq:Sackurnorm}) gives
\begin{equation}
\frac{G}{\tilde{N}T}=k \left( \frac{3}{2}\ln \frac{2 \pi m}{h^2 \beta}-\ln \eta \right),
\label{eq:phiideal}\end{equation}
so that the PTN ensemble becomes
\begin{equation}
p_{_{\rm {PTN_{id}}}}(\hat{u},\hat{v})=\mathcal{Z} \hat{u}^{3/2}e^{-\beta \hat{u}}\hat{v} e^{-\eta \hat{v}},
\end{equation}
with $\hat{u}=\hat{U}/\tilde{N}$, $\hat{v}=\hat{V}/\tilde{N}$, and $\mathcal{Z}$ the normalization constant. The distribution is the product of two gamma distributions, which underlines the statistical independence of $\hat{u}$ and $\hat{v}$ for the case of the ideal gas.

For constant volume, using again Eq.\ (\ref{eq:Sackurnorm}), the canonical ensemble for ideal gas is
\begin{equation}
p_{_{\rm {TVN_{id}}}}(\hat{u};\beta,V,N)=\Omega_0 (2/3e)^{3/2}(\beta \hat{u})^{3/2}e^{-\beta\hat{u}}. 
\end{equation}
Upon normalization with
\begin{equation}
    \Omega_0=\frac{4}{3\pi}(2/3e)^{-3/2}\beta,
\end{equation}
one obtains a gamma distribution for the specific internal energy
\begin{equation}
p(\hat{u})=\frac{4}{3\pi}\beta^{5/2} \hat{u}^{3/2}e^{-\beta \hat{u}}
\end{equation}
with mean $\langle \hat{u}\rangle=\frac{5}{2 \beta}=5/3 u$, mode $\breve{u}=\frac{3}{2 \beta}=u$, and variance ${\rm var}(\hat{u})=5/2\beta^{-2}$. Because of the normalization, the variance is larger than the one 
obtained from considering $\beta F$ as the cumulant transform, $c_v T^2=3/2\beta^{-2}$ \cite{callen1998thermodynamics,pathria2016statistical}. This point deserves further investigation. Here we only note that the reason might be related to the saddle-point approximation typical of the Gibbs ensembles \cite{pathria2016statistical,touchette2009large}, which would imply that $\Phi$ is not exactly the cumulant transform of the distribution
\cite{barndorff1989asymptotic,porporato2014dual}. In statistics, normalization of the distribution has been shown to lead to better inferences than the saddle point approximation \cite{daniels1980exact,reid1988saddlepoint}, while in large deviations \cite{touchette2009large} the Legendre transform of the rate function, by itself, 
gives un-normalized approximations: it only gives their asymptotic shape. 

\begin{figure}
\begin{center}
\includegraphics[width=250pt]{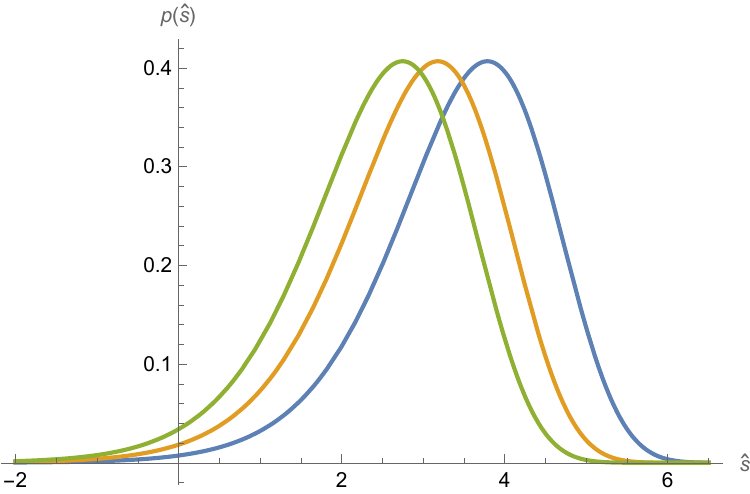}
\end{center}
\caption{Pdf of the particle specific entropy for a monoatomic ideal gas (helium) for three different equilibrium inverse temperatures ($\beta=0.2,0.3,0.4$).} \label{fig:pdfentr}
\end{figure}

The analytical form of the canonical distribution allows us also to obtain the microcanonical specific entropy distributions as
\begin{equation}
    p_{_{\hat{s}}}(\hat{s})=\mathcal{N}e^{\frac{5}{3}\hat{s}/k-\beta e^{\frac{2}{3}\hat{s}/k}},
\end{equation}
which is reminiscent of the extreme value I (or Gumbel) distribution (for its negative) and has mean proportional to $\ln \beta$ and constant variance (see Fig. 3). The previous explicit distributions make the extrapolation underlying the ensembles particularly evident. In particular, the ideal gas model, which is valid only at high temperatures (the Sackur-Tetrode entropy (\ref{eq:Sackurnorm}) even becomes negative below a certain temperature), when used as a fundamental equation in the exponent of the ensemble opens up an infinite range for internal energy, $0\le\hat{u} < \infty$. One obviously needs to handle the regions of the ensemble away from the bulk of the distribution with great attention, as it is unlikely that simply adapting the fundamental equation to the whole range of fluctuations explored, including phase changes, might produce physically sensible thermodynamic fluctuations. 

\section{Stochastic Heat Transfer}

We consider here a condition of heat transfer obtained from the TLEs for the canonical ensemble in the presence of probability currents. We focus again on the case of a monoatomic ideal gas, for which the equation for the specific internal energy reads
 \begin{equation}
     d\hat{u}=\left(\frac{1}{T}-\frac{2}{3\hat{u}}\right) dt+d\hat{W},
     \label{eq:TLEideal}
 \end{equation}
with $\alpha, k=1$ for simplicity. In principle, a heat transfer can be obtained by imposing a probability current between two values of internal energy, $\hat{u}_1$ and $\hat{u}_2$. This is achieved by opening the boundaries and (say) removing some trajectories at $\hat{u}_2$, while re-injecting them at $\hat{u}_1$. Thus a given heat flux (a constant probability current, $J_u$) is produced by adding trajectories with low energy at $\hat{u}_1$ and 'killing' them at $\hat{u}_2>\hat{u}_1$. Physically, this may be associated to the case of a homogeneous gas, in equilibrium with a heat bath at temperature $T$, in which some fraction of particles are individually and instantaneously cooled to (or removed and substituted with a particle at) a predetermined lower energy $\hat{u}_1$, once their fluctuating energy reaches a certain maximum level of energy $\hat{u}_2$ (which is analogous to a molecular dynamics algorithm for shearing flows \cite{muller1999reversing}). 

The ensuing non equilibrium steady state (NESS) pdf can be obtained analytically and written implicitly in 'potential form' as \cite{porporato2011local}
\begin{equation}
    p_{J}(\hat{u})=\mathcal{N}e^{-\Sigma(\hat{u})-\int_{\hat{u}_1}^{\hat{u}} J_u/p_{_{J}}(x') dx'}.
    \label{eq:solcurr}
\end{equation}
Interestingly, the presence of the extra potential due to the current means that, formally, one can invert the procedure which leads to (\ref{eq:enspot}) to obtain a fundamental equation having the current as one of the macrostate variables. Analogous expressions for the pdf have been suggested in \cite{miller1979heat,kato2001breaking,conti2013effects}, albeit with no relation to TLEs. 

\begin{figure}
\begin{center}
\includegraphics[width=250pt]{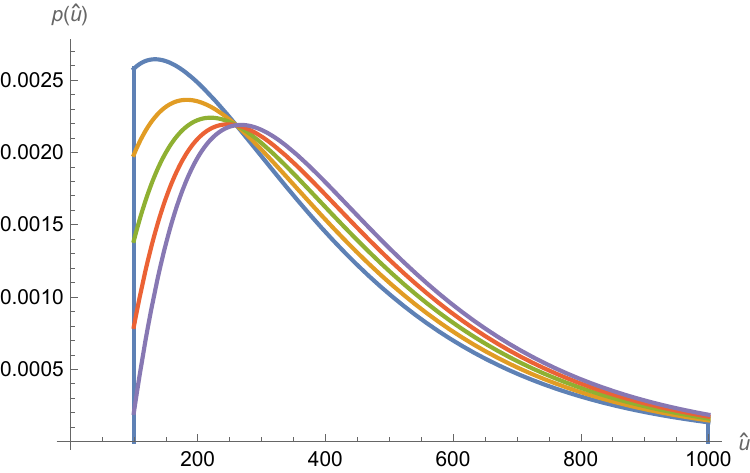}
\end{center}
\caption{Solution (\ref{eq:solcurr}) of the Langevin equation (\ref{eq:TLEideal}) with different probability currents. Blue curve is the renormalized canonical distribution confined between two fixed values, $\hat{u}_1$ and $\hat{u}_2$, with $\hat{u}_1<\hat{u}_2$ functioning as reflecting boundaries. Other curves with lower maxima refer to modifications of the pdf due to positive probability currents of increasing strength. At the maximum value of the probability current the lower end of the pdf becomes zero.} \label{fig:canJ}
\end{figure}

The specific distributions for increasing values of the current are shown in Fig. \ref{fig:canJ}. The current increases from zero, in case of reflecting boundaries, to a maximum allowable value, where the pdf becomes zero at $\hat{u}_1$. It is logical to inquire whether an actual thermodynamics of heat transfer emerges from this formal extension of the canonical ensemble. As a reference, we consider the NESS heat transfer between two fixed temperatures. According to classical thermodynamics, the entropy balance is the difference between the entropy outflow and inflow, namely
\begin{equation}
    J_q\left(\frac{1}{T_1}-\frac{1}{T_2}\right),
\label{eq:entrprodclass}\end{equation}
where $J_q$ is the rate of heat flux. For a monoatomic ideal gas,  $T_1=\frac{2}{3k}u_1$ and
$T_2=\frac{2}{3k}u_2$. This expression (\ref{eq:entrprodclass}) should be compared with the rate obtained with the stochastic thermodynamics approach \cite{qian2001mesoscopic,seifert2005entropy,porporato2011local, seifert2012stochastic}, according to which the mesoscopic (i.e., at a point) entropy balance for a Langevin equation with drift $a$ and unitary additive noise reads
\begin{equation}
    \frac{ds}{dt}=\frac{d\ln p}{dt}=-\frac{\partial p}{\partial t}-2va+\sigma,
\end{equation}
where $\sigma=2 v^2$ is the local entropy-generation rate. As a result, for the steady state case with constant current,
\begin{equation}
    \int_{\hat{u}_1}^{\hat{u}_2}  \sigma(\hat{u}) p(\hat{u}) d \hat{u}=2 J_u^*\int_{\hat{u}_1}^{\hat{u}_2}  \left(\frac{1}{T}-\frac{2}{3\hat{u}}\right)d\hat{u}.
    \label{eq:enpot1}
\end{equation}
The solution, expressed in terms of temperature, $\hat{T}=\frac{2}{3}\hat{u}$, becomes
\begin{equation}
    2J_u^*\left(\frac{\hat{T}_2-\hat{T}_1}{T}-\ln\frac{\hat{T}_2}{\hat{T}_1} \right).\label{eq:enpot2}
\end{equation}
The expressions (\ref{eq:enpot1}) and (\ref{eq:enpot2}) represent two different heat transfer conditions. To find a physical process corresponding to the NESS solution of the Langevin equation (\ref{eq:TLEideal}) as an actual heat-transfer process is perhaps necessary to remain in the realm of small systems, as averages of single particle experiments, which do not have a macroscopic counterpart. In addition, the fact that for the stochastic process there is a maximum current, beyond which the pdf becomes negative (see Fig. 4), corresponds to a maximum of rate of heat flux and entropy generation, while the system always remains in contact with the heat bath. 

\section{Isothermal Colloidal Particle: Stochastic Dynamics of Micro- and Macro-states}\label{sec:collo}

It is instructive to compare the TLEs with the equations used in stochastic thermodynamics, which considers a colloidal particle confined to one spatial dimension as the paradigm for the field \cite{seifert2012stochastic}. For simplicity we assume the force to be constant and acting along one of the axes. This happens for example when a colloidal 
particle is pulled by an optical tweezer, or when a neutrally buoyant ferromagnetic particle, immersed in a fluid, is driven by a constant magnetic field. It is also the case of a very small particle settling in the gravitational field in the presence of thermal fluctuations.

We indicate with $0 \le l \le L_0$ the confined particle position in the direction of the force, while in the other two directions the particle density is uniform. 
The Hamiltonian is $\mathcal{H} = 1/2 m {\bf v}^2 + f l$. In turn, the marginal pdf for the particle position is easily obtained as a truncated exponential distribution,
\begin{equation}
    p(\hat{l})=\int_{{\bf v}} Z e^{-\beta \mathcal{H}}=Z_{l}e^{-\beta f \hat{l}}
    \label{eq:collomicro}
\end{equation}
for  $0 \le l \le L_0$. 
This is a condition of equilibrium, where the particle distribution is made inhomogeneous by the external field. 

\begin{figure*}
\begin{center}
\includegraphics[width= 0.8 \textwidth]{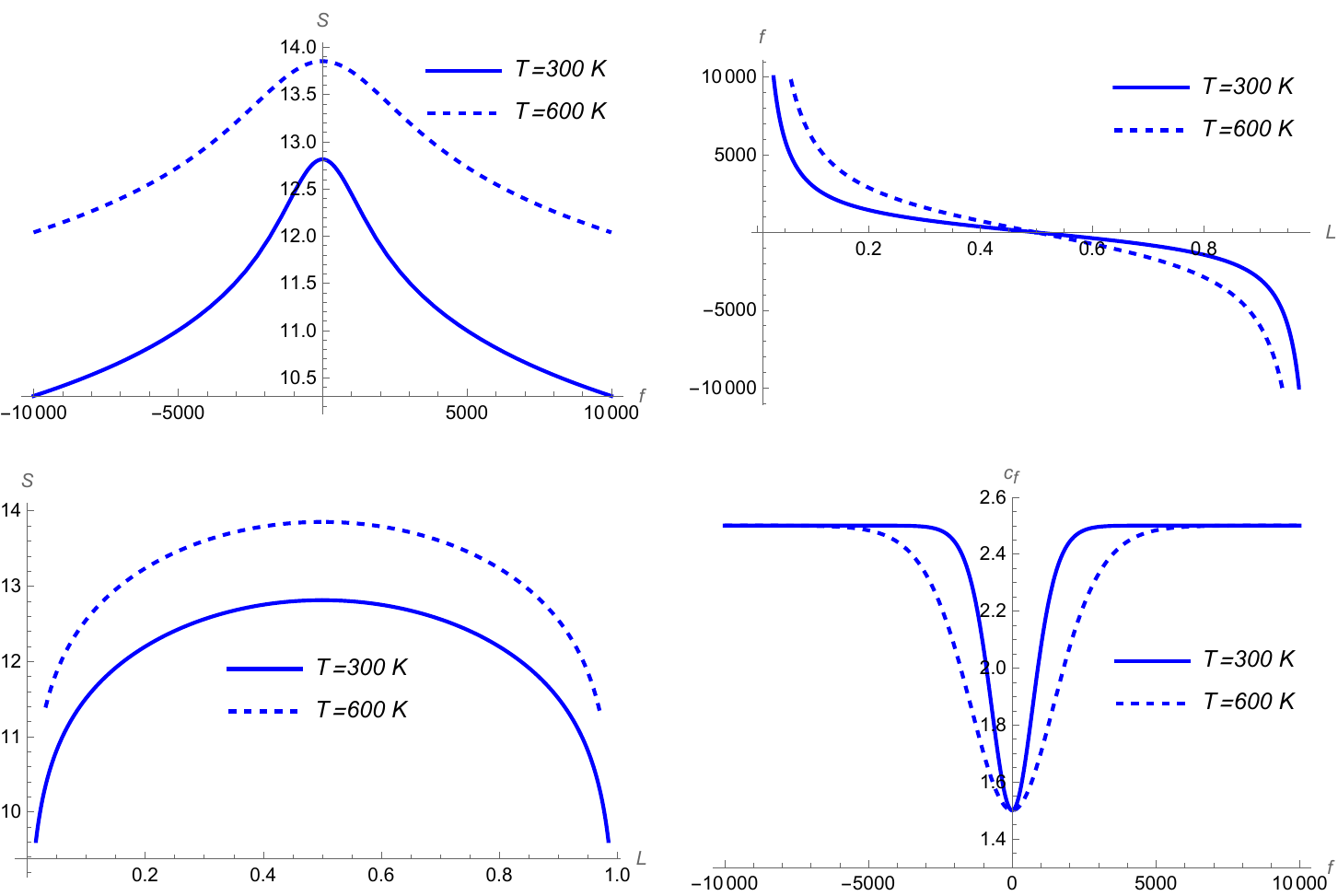}
\end{center}
\caption{Small system thermodynamics of colloidal particle under constant force. (Top left) Entropy as a function of force for different temperatures, showing the ordering effect caused by the interaction; (Top right) Equation of state $f(U, L)$ plotted as a function of $L$. (Bottom left) Entropy as a function of $L$; (Bottom right) Isocoric heat capacity as a function of $f$. All graphs are drawn for $L_0 = 1$, $k = 1$, and $m=1$.} \label{fig:collo1}
\end{figure*}

The macroscopic thermodynamic quantities are obtained from the partition function 
\begin{equation}
    Z=e^{-\beta F_e}=\int_0^{L_0}dl\int_{-\infty}^{\infty} d{\bf v} e^{-\beta \left(\frac{m{\bf v}^2}{2}+fl\right)},
\end{equation}
through its link with the extended free energy $F_e(T,f)=U+fL-ST$. Here, $L$ is the average position of the particle in the direction of the force. The entropy can be obtained as
\begin{equation}
    S=-\frac{\partial F_e}{\partial T}=k\ln\left(\frac{2\sqrt{2} \pi^{3/2}\psi}{f\left(\frac{1}{kT}\right)^{5/2}} \right)+\frac{f L_0}{T\psi}+\frac{5k}{2},
\label{eq:entpart}\end{equation}
where $\psi=1-e^{\frac{fL_0}{kT}}$, while
\begin{equation}
    L=\frac{\partial F_e}{\partial f}=\frac{kT}{f}-\frac{L_0}{e^{\frac{kT}{f L_0}-1}} 
\label{eq:lpart}\end{equation}
and the internal energy is 
\begin{equation}
    U=F^e-fL+ST=\frac{3}{2}kT.
\label{eq:upart}
\end{equation}
Combining the entropy (\ref{eq:entpart}) with equations (\ref{eq:lpart}) and (\ref{eq:upart}) yields the fundamental equation, $S = S(U,L)$. This is a monotonically increasing function of the energy $U$ (as well as temperature $T$); however, when plotted as a function of $L$ (or $f$), it has a
maximum at $L = L_0/2$ (or $f=0$) and decreases as $L$ tends to zero or 1 (or correspondingly $f$ decreases or increases from zero) -- see Fig. \ref{fig:collo1}). This symmetrical bell shape
originates from the 
ordering effect brought about by the conservative
interaction with the environment, which forces the particle to spend more time in some positions rather than others \cite{calabrese2019origin}. Thus, while in the absence of the field the particle uniformly explores the entire volume, and the entropy is at its maximum, when the interaction is turned on, the particle is pushed towards the wall and the entropy is reduced. The interaction with the environment also affects the material properties of the system.
The heat capacity, $c_f=T\frac{\partial^2 F^e}{\partial T^2}|_f$ is $3/2k$ at $f = 0$ and increases as the interaction becomes stronger (Fig. \ref{fig:collo1}, bottom right). 

\begin{figure}
\begin{center}
\includegraphics[width=250pt]{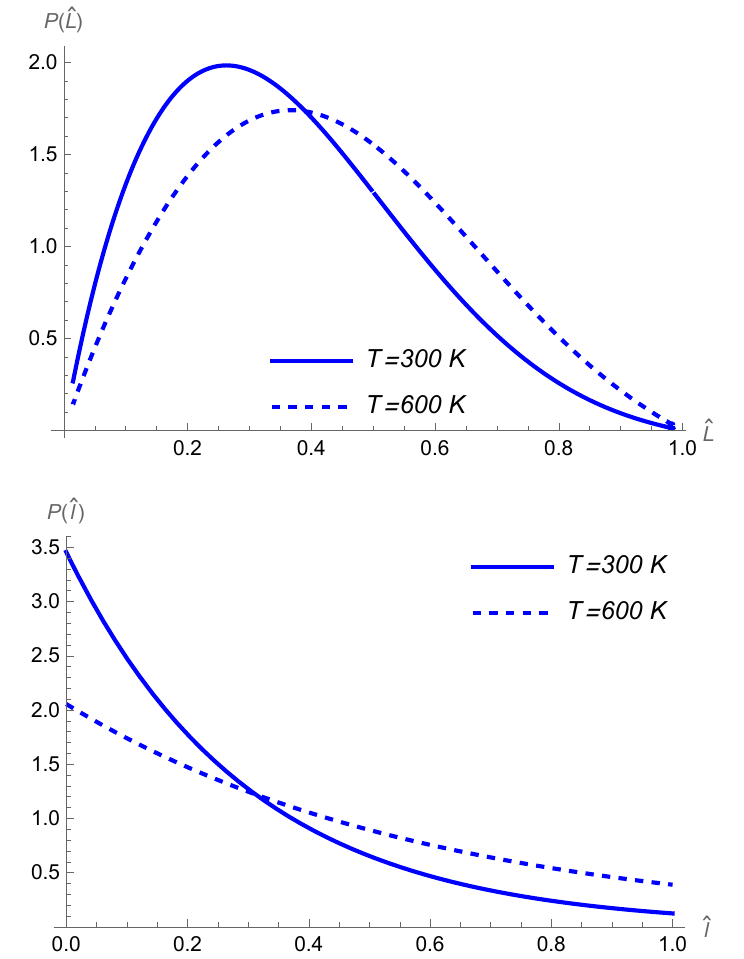}
\end{center}
\caption{The conditional distribution for $\hat{L}$ obtained from (\ref{eq:enscollo}) for a given value of $\hat{U}$ (that is corresponding to a given temperature), shown for two values of temperature (top). Exponential distribution for the microscopic quantity $\hat{l}$ (bottom). Graphs drawn for $L_0 = 1$, $k = 1$, and $m=1$.} \label{fig:collo2}
\end{figure}

The thermodynamic fluctuations of $U$ and $L$ are described by the bivariate TfN-ensemble 
\begin{equation}
    p_{_{\rm TfN}}(\hat{U},\hat{L})=\Omega_0 e^{-\beta (F_e -\hat{U}+f\hat{L}+T\hat{S})}
    \label{eq:enscollo}
\end{equation}
and the TLEs ($\alpha,k=1$)
\begin{eqnarray}
    d \hat{U}=\left( \frac{1}{T}-\frac{\partial \hat{S}}{\partial \hat{U}}\right)dt+d\hat{W}_1 
    \label{eq:collostoch1}\\
    d \hat{L}=\left( \frac{f}{T}-\frac{\partial \hat{S}}{\partial \hat{L}}\right)dt+d\hat{W}_2. \label{eq:collostoch2}
\end{eqnarray}
The equilibrium distribution, obtained with natural boundary conditions \cite{gardiner1985handbook}, gives (\ref{eq:enscollo}) by design. The actual expression is involved due to some implicit functions, which apparently make $\hat{U}$ and $\hat{L}$ statistically dependent. The conditional distribution for a given value of $\hat{U}$ is shown for two values of temperature in Fig. \ref{fig:collo2}, where it is compared with the exponential distribution for the microscopic quantity $\hat{l}$.  

Rather than using (\ref{eq:collostoch1}) and (\ref{eq:collostoch2}), stochastic thermodynamics \cite{seifert2005entropy} uses the microscopic quantity $\hat{l}$ as the fluctuating thermodynamic quantity. As before for the macrostates, e.g., Eq. (\ref{eq:enspot}), the position distribution (\ref{eq:collomicro}) is in potential form, but now this is only a special condition due to the 1D symmetry, not a general fact, which allows (\ref{eq:collomicro}) to be interpreted as the solution of the Langevin equation
\begin{equation}
    d \hat{l}=-f+ k T d\hat{W}_l
    \label{eq:collostochlang}
\end{equation}
with reflecting boundary conditions at $\hat{l}=0$ and $L_0$.

Stochastic thermodynamics adopts the stochastic energetics of Sekimoto \cite{sekimoto1998langevin}. Accordingly, for equilibrium conditions, the changes in potential energy, due to changes of position in the field, are considered as (reversible) changes in internal energy, $du$, which in turn are associated with heat taken up by the reservoir, $dq^{{\rm rev}}$. 
The potential responsible for the force is thus considered as internal to the system, instead of being considered as an external field as in (\ref{eq:collostoch1}) and (\ref{eq:collostoch2}); see also \cite{calabrese2019origin}. The entropy dependence on the spatial variable is also different: since in equilibrium thermodynamics $dq^{{\rm rev}}=Tds$,
$du=fd\hat{l}=Tds$, which implies $s\propto\hat{l}$, in agreement with Seifert's stochastic entropy based on the logarithm of the distribution (\ref{eq:collomicro}).
This would also result if one interprets the SDE (\ref{eq:collostochlang}) as a TLE (\ref{eq:TLE}), 
or equivalently assuming that the distribution for the microstates (\ref{eq:collomicro}) is indeed a distribution for the macrostates (\ref{eq:ens}); this leads to a thermodynamics with constant entropic forces in $\hat{l}$, which in turn originate from an entropy linear in $\hat{l}$. Further differences likely would be found in the presence of probability currents, but this will be explored elsewhere.

\section{Conclusions}

Thermodynamics great achievement is to describe macroscopic system with few sufficient statistics \cite{mandelbrot1962role} that allow neglecting the noisy mess of the myriad of dynamic degrees
of freedom. For small system, this is no longer possible and one's hope is to be able to find a physically meaningful description without too many sacrifices.

Stochastic thermodynamics \cite{seifert2012stochastic}, which has flourished recently, directly starts from stochastic versions of the first and second law of thermodynamics and introduces fluctuating thermodynamic quantities building on microscopic variables. In this paper we followed a different approach by considering stochastic variability directly in the macroscopic variables. By recognizing the potential structure of the Gibbs ensembles, when expressed as a function of the potential entropy generation, 
$\Sigma$, we have obtained stochastic differential equations in terms of entropic forces. The resulting TLEs by construction exactly reproduce the generalized Gibbs ensembles in equilibrium and are in line with the classical theory of thermodynamic fluctuations and their underlying structure \cite{callen1998thermodynamics,landau2013statistical}. In comparison, stochastic thermodynamics \cite{seifert2012stochastic} although having consistent first and second laws, has not yet been embedded in a complete thermodynamic edifice, which includes fundamental equations, equations of state and material properties for the small system in equilibrium.

The real challenge of a small system theory resides in the nonequilibrium realm. In the presence of probability currents, the canonical entropy obtained from the TLEs acquires a production term as the macroscopic variable undergo transient or NESS protocols. The results obtained for a stochastic heat transfer reveal interesting bounds on the admissible currents and make explicit the ever-present role of the environment to which the nonequilibrium system remains coupled. As a consequence, the resulting nonequilibrium systems do not have immediate physical interpretation in term of classical nonequilibrium thermodynamics. 

Our results also point to the ways the generalized ensembles morph in nonequilibrium conditions, at the same time suggesting caution for too literal interpretations of such extensions; this is especially true for results involving parts of the distributions (i.e. the tails) which are too far away from the central region for which their asymptotic forms are originally conceived. We hope that future work will help clarify some of these outstanding points as well as find common ground among the different approaches to small systems.

\bibliography{references}
\end{document}